\documentclass[prd,nofootinbib,showpacs,twocolumn,reprint]{revtex4}
\usepackage[a4paper, total={6.5in, 8.5in}]{geometry}
\usepackage{graphicx}
\usepackage{bm}
\usepackage[usenames,dvipsnames]{color}
\usepackage{amsmath,amssymb}
\usepackage{slashed}
\usepackage{float}
\usepackage{epsfig}
\usepackage{latexsym, amssymb} 
\usepackage[colorlinks=true,linkcolor=blue,citecolor=blue]{hyperref}
\begin{document}
	
\title{Formation of Primordial Black Holes from Warm Inflation} 	
\author{Richa Arya}
\email{richaarya@prl.res.in}
\affiliation{
Theoretical Physics Division, Physical Research Laboratory, Navrangpura, Ahmedabad 380009, India}
\affiliation{
Indian Institute of Technology Gandhinagar,
Palaj, Gandhinagar 382355, India}

\date{\today}

\begin{abstract}
Primordial Black Holes (PBHs) serve as a unique probe to the physics of the early Universe, particularly inflation.
In light of this, 
we study the formation of  
PBHs by the collapse of  overdense perturbations 
generated during a model of warm inflation. For our model, we find that
the primordial curvature power spectrum 
is red-tilted (spectral index $n_s<1$) at the large scale (small $k$) and is consistent with the $n_s-r$ values allowed from the
CMB 
observations.
Along with that, it has a blue-tilt ($n_s>1$) for the small PBH scales (large $k$), with a sufficiently large amplitude of the primordial curvature power spectrum required to form PBHs.
These features originate beacause of the inflaton's coupling with the other fields during warm inflation.
We discuss the role of the inflaton dissipation 
to the enhancement in 
the primordial power spectrum at the PBH scales. 
We find that for some parameter range of our warm inflation model,  PBHs with mass $\sim 10^3$ g can be formed with significant abundance. Such tiny mass PBHs have a short lifetime $\sim 10^{-19}$ s and would have evaporated into Hawking radiation in the early Universe. Further in this study, we discuss the evaporation constraints on the initial mass fraction of the generated PBHs and the possibility of Planck mass PBH relics to constitute the dark matter. 
\end{abstract}

\maketitle
\newpage

\section{Introduction}
Primordial Black Holes (PBHs) \cite{zeldo:1966,Hawking:1971ei,Carr:1974nx, Carr:1975qj} are the black holes that could have produced in the very early Universe. 
They can form in a number of ways - by the collapse of overdensities in the early inhomogeneous Universe \cite{Hawking:1971ei,Carr:1974nx}, from the collision of bubbles \cite{Hawking:1982ga}, by the collapse of strings \cite{Hogan:1984zb,Hawking:1987bn} or domain walls \cite{Caldwell:1996pt}, etc. 
For a review on PBHs, see Refs. \cite{Carr:2005zd,Khlopov:2008qy,Green:2014faa,Sasaki:2018dmp}.

It is crucial to study primordial black holes as they provide us a unique probe to the rich physics of the Universe at all epochs of its evolution.
While the Cosmic Microwave Background (CMB) and Large Scale Structures (LSS) observations measure only the large scale modes ranging from $10^{-3}-1$ Mpc$^{-1}$, PBHs span over a wide range of modes varying from $10^{-2}-10^{23}$ Mpc$^{-1}$, and therefore provide a probe for a huge range of small scales.  
The mass of a PBH depends on the epoch when it is formed, with lighter mass PBHs forming earlier than the heavier PBHs. 
The lightest PBH mass 
can be as tiny as
 corresponding to the Planck mass, $M_P=10^{-5}$ g \cite{Carr:1974nx}. 
 
 The abundance of PBHs 
 is constrained from various observations. These constraints can be classified for two mass ranges of PBHs, $M_{PBH}>10^{15}$ g, which are obtained from the gravitational effects of PBHs, 
 and $M_{PBH}\leq 10^{15}$ g, obtained from the effects of their evaporation.
 PBHs with mass $M_{PBH}<10^{15}$ g have short lifetime compared to the present age of the Universe, and would have evaporated into  Hawking radiation by the present time \cite{Hawking:1974rv, Hawking:1974sw}. 
Therefore, for PBHs with $M_{PBH}<10^{15}$ g, the consequences of PBH evaporation on the nucleosynthesis \cite{vain, zeldo,Kohri:1999ex}, or the relic abundance of stable and long lived decaying particles produced from PBH evaporation \cite{ Green:1999yh,Lemoine:2000sq,Khlopov:2004tn} can provide constraints on their abundance (see Ref. \cite{Carr:2009jm}).  The upper limit on PBH abundance further give bounds on the amplitude of the primordial curvature power spectrum and hence various inflationary models. (For details, see Refs. \cite{Carr:1993aq,Kim:1996hr,Green:1997sz,Josan:2009qn}). In this way, PBHs can serve as a powerful and unique probe to the inflationary epoch. 

Furthermore, PBHs with $M_{PBH}\sim10^{15}$ g would be evaporating into radiation at the present epoch and have interesting astrophysical consequences.
Such PBHs can contribute to the diffuse gamma-ray background \cite{Page}, or positrons and antiprotons in the cosmic rays  \cite{MacGibbon:1991vc} and therefore can 
provide useful information about the 
high energy 
physics of the PBH evaporation \cite{Carr:2009jm}.

The primordial black holes  with $M_{PBH}>10^{15}$ g have not evaporated till today.
An interesting consequence of such PBHs is that 
they can 
contribute as some or all of the Dark Matter (DM) present in the Universe, and therefore their abundance should be less than the limits on the cold dark matter density at present.   
(For review, see Ref. \cite{Carr:2016drx,Ballesteros:2017fsr}). 
The constraints on the abundance of these PBHs 
can also be obtained
from the different lensing experiments \cite{Paczynski:1985jf, Gould, Hawkins}, or from the dynamical effects of PBHs on the astrophysical systems (For a review, see Ref. \cite{Sasaki:2018dmp}). 
Also, massive PBHs (few solar masses $M_\odot$) can 
accrete its surrounding gas
and emit  X-rays, which can change the ionization history of the Universe \cite{10.1093/mnras/194.3.639} and cause spectral distortions in the CMB radiation \cite{Ricotti:2007au, Ali-Haimoud:2016mbv}. Therefore, the abundance of massive PBHs is constrained from the CMB anisotropy observations. 
Recent observations of microlensing of the stars in the Andromeda galaxy suggest that PBHs in the mass range $10^{19}$ g $<M_{PBH}<10^{24}$ g can constitute only a small fraction $<0.1\%$ of the dark matter energy density \cite{Niikura:2017zjd}.  However, 
the possibility that PBHs comprise a significant fraction of DM still remains open 
for other mass ranges of PBHs,
like 
 $M_{PBH}\sim (10-100) M_{\odot}$ 
detected in the gravitational wave signals  
by the LIGO collaboration 
 \cite{PhysRevLett.116.201301,CLESSE2017142}. (For review, see Refs. \cite{Garcia-Bellido:2017fdg,Sasaki:2018dmp}.)   
It is also argued that
 Planck mass ($M_P=10^{-5}$ g) remnants of the evaporating PBHs can comprise the dark matter \cite{MacGibbon:1987my,Chen:2003bn}. We will discuss it further in our study.

In this work, we study an inflationary scenario known as
Warm Inflation \cite{Berera:1995wh,Berera:1995ie,Berera:1998px} and discuss the formation of primordial black holes by the collapse of large inhomogeneities generated during it. Warm Inflation is a description of inflation in which the inflaton is coupled sufficiently enough to the other fields 
and dissipates its energy into them even during inflation. 
As a result of the inflaton's coupling and dissipation, the Universe constitutes a thermal bath of particles and has a temperature throughout the inflation. For a review on Warm Inflation, see Refs. \cite{Berera:2006xq,Berera:2008ar}. The primordial power spectrum for warm inflation is dominated by the contributions from the thermal fluctuations of the fields, unlike the quantum fluctuations in cold inflation, as will be explained in Section \ref{WI}.   
In this paper, we discuss the features in the primordial power spectrum of a model of warm inflation for various values of the dissipation parameter and then study the formation of PBHs from our model.  

We find that for our warm inflation model, the primordial power spectrum is red-tilted (spectral index, $n_s<1$) for the CMB scales, with an amplitude $P_{\mathcal R}(k_P)=2.1\times 10^{-9}$ (at the pivot scale $k_P=0.05$ Mpc$^{-1}$), and is consistent with the $n_s$ and $r$ (tensor-to-scalar ratio)  values allowed from the CMB observations. Apart from this, it has a blue-tilt ($n_s>1$) with a large amplitude
 of the primordial power spectrum
for the PBH scales.
In our analysis, we find that 
for some range of the model parameter,
the amplitude of $P_\mathcal{R}(k)$ at the PBH scales is of $\mathcal{O}(10^{-2})$ which is required for the PBH formation, and
therefore, a significant abundance of PBHs can be produced. 
We first discuss the relevant range of model parameter and then calculate the initial mass fraction and the mass of the generated PBHs.

As mentioned previously, PBHs are a unique probe to  inflation, as the observational bounds on the abundance of PBHs provide an upper limit on the amplitude of the primordial power spectrum. Therefore, a study of PBHs is crucial to test various inflationary models. 
Here we carry out the first study considering a warm inflation primordial power spectrum for the PBH formation.  The importance of this work is that our warm inflation model is also consistent with the CMB bounds on  $n_s-r$  and the theoretical prediction for the tensor-to-scalar ratio is within the sensitivity of the upcoming CMB polarisation experiments and hence can be tested in the near future.
 
The organization of this work is as follows.  In Section \ref{PBH form}, we first discuss the formation of primordial black holes in terms of the mass of the generated PBH and its initial mass fraction. 
Then using the Press-Schechter formalism, we calculate the 
bounds on the primordial curvature power spectrum from the bounds on PBH abundance. 
After that, we discuss the fundamentals of warm inflation and the primordial power spectrum in Section \ref{WI} .
 Then we present our model of warm inflation in Section \ref{ourWI} and analyze the obtained results in Section \ref{discussion}. At last we conclude our study in Section \ref{Summary}.

\section{Formation of a Primordial Black Hole}
\label{PBH form}
As mentioned previously, PBHs can be generated through many mechanisms. Here we consider 
the PBH formation by the collapse of overdense perturbations generated during a model of warm inflation. These fluctuations leave the horizon during inflation and then re-enter at later epochs (we assume radiation dominated era), and collapse to form PBHs. In this section, we first calculate
the mass of PBHs formed as a function of the fluctuation mode, $k$.
 We then define the initial mass fraction of PBH and discuss the constraints on  it 
 from observations. Further, we discuss the Press-Schechter formalism for PBH formation and using it, 
 calculate the initial mass fraction of PBHs. 
After that, we show how the  constraints on the abundance of PBHs can be used to obtain bounds on the amplitude of the primordial power spectrum.

\subsection{Mass of the generated Primordial Black Holes} 
\label{mass}

Primordial black hole forms when an overdense fluctuation with a comoving wavenumber $k$ generated during  inflation, re-enters the horizon at a later epoch (i.e. physical wavelength equals the horizon size,  $H^{-1}=(k/a)^{-1}$ 
) with an overdensity $\delta$ greater than a critical density $\delta_c$, and collapses through gravitational instability.
The mass of the generated PBH, $M_{PBH}$ depends on the time of its formation and is taken to be a fixed fraction, $\gamma$,  of the horizon mass at that epoch \cite{Green:1997sz}, 
\begin{equation}
	M_{PBH}(k)=\gamma \frac{4\pi}{3}\rho \left.H^{-3}\right|_{k=aH}.
	\label{MPBH}
\end{equation}
Here $H$ is the Hubble expansion rate and $\rho$ is the energy density of the Universe at the epoch of PBH formation.
 We consider that the PBH formation takes place during the radiation dominated era, and therefore $\rho$ is the energy density of the radiation, i.e. $\rho=\rho_r$. The fraction of the horizon mass collapsing into the PBHs is taken as $\gamma=0.2$ \cite{Carr:1975qj}. 

In order to calculate the relation between the $k^{th}$ mode of perturbation, and the mass of the generated PBH, we write the parameters on the r.h.s. of Eq. (\ref{MPBH}) explicitly as a function of  $k$. This is carried out  as follows.
 
 The energy density of the radiation is given as $\rho_r=\frac{\pi^2}{30}g_* T^4,$ where $g_*$ is the number of relativistic degrees of freedom, and $T$ represents the temperature of the Universe in the radiation dominated era.
 Following the conservation of entropy, we have $S=g_{*s}a^3T^3=$ constant, where $S$ is the entropy, $a$ is the scale factor of the Universe, and  $g_{*s}$ represent the number of relativistic degrees of freedom contributing in the entropy. We assume that the number of relativistic degrees of freedom contributing to the radiation equals to the ones contributing in the entropy i.e. $g_* \approx g_{*s}$, and thus obtain $\rho_r\propto g_*^{-1/3}a^{-4}.$ With this relation, the radiation energy density at the initial time of PBH formation $\rho_{r i}$ can be related to the present radiation energy density $\rho_{r0}$ as
 \begin{equation}
 \rho_{ri}= \left(\frac{g_{*i}}{g_{*0}}\right)^{-1/3}\left(\frac{a_i}{a_0}\right)^{-4}\rho_{r0}.
 \label{rhori}
 \end{equation}
 The subscript `$0$' and `$i$' to any quantity represent its value at the present epoch and at the initial time when PBH formed, respectively.  The scale factor at present $a_0=1$.
The present radiation density can be written as $\rho_{r0}=\rho_{crit}~\Omega_{r0}$, where $\rho_{crit}=3 H_0^2 /8\pi G_N =1.054\times10^{-5} h^2$ GeV cm$^{-3}$ is the the critical energy density today, $G_N$ is the Newton's gravitational constant, $H_0=100~ h$ is the present Hubble expansion rate with $h=0.678$, and $\Omega_{r0} \approx5.38 \times 10^{-5}$ is the radiation density parameter today \cite{Tanabashi:2018oca}. 
Thus we can write Eq. (\ref{rhori}) as
\begin{equation}
\rho_{ri}= \left(\frac{g_{*i}}{g_{*0}}\right)^{-1/3}a_i^{-4}\rho_{crit}~\Omega_{r0}~,
\label{rhoi}
\end{equation}
and substitute it in Eq. (\ref{MPBH}) to obtain the mass of the generated PBH as 
\begin{align}
M_{PBH}(k)=&\gamma \frac{4\pi}{3}\left(\frac{g_{*i}}{g_{*0}}\right)^{-1/3}a_i^{-4}\rho_{crit}~\Omega_{r0}\left(\frac{k}{a_i}\right)^{-3} \nonumber\\
=&\gamma \frac{4\pi}{3}\left(\frac{g_{*i}}{g_{*0}}\right)^{-1/3}a_i^{-1}\rho_{crit}~\Omega_{r0}~k^{-3}.
\label{MPBH2}
\end{align} 

Now we will determine the scale factor  $a_i$ at the time of PBH formation. 
The Hubble rate of expansion at the time of PBH formation, when $k^{th}$ fluctuation mode re-enters the horizon can be written as,
\begin{equation}
H^2=\left(\frac{k}{a_i}\right)^2=\frac{8\pi G_N}{3}\rho_{ri}.
\end{equation} 
By substituting the expression for the initial radiation energy density from Eq. (\ref{rhoi}) into this, we get
\begin{equation}
\left(\frac{k}{a_i}\right)^2=\frac{8\pi G_N}{3}\rho_{crit}\left(\frac{g_{*i}}{g_{*0}}\right)^{-1/3}a_i^{-4}\Omega_{r0}
\end{equation}
which gives 
\begin{equation}
a_i^{-1}=\left(k^2 H_0^{-2} \left(\frac{g_{*i}}{g_{*0}}\right)^{1/3}\Omega_{r0}^{-1}\right)^{1/2}.
\label{ai}
\end{equation}
Finally, we substitute Eq. (\ref{ai}) in Eq. (\ref{MPBH2}) and obtain
\begin{align}
M_{PBH}(k)=&\gamma \frac{4\pi}{3}\left(\frac{g_{*i}}{g_{*0}}\right)^{-1/3} k H_0^{-1}\left(\frac{g_{*i}}{g_{*0}}\right)^{1/6} \nonumber \\ & \times \Omega_{r0}^{-1/2}~  \rho_{crit}~\Omega_{r0}~k^{-3} \nonumber\\
=&\gamma \frac{4\pi}{3}\rho_{crit}\left(\frac{g_{*i}}{g_{*0}}\right)^{-1/6}~\Omega_{r0}^{1/2} ~k^{-2}H_0^{-1}. 
\label{MBH3}
\end{align}
This relation implies that the mass of the generated PBH is proportional to the inverse square of the $k^{th}$ mode of fluctuation that creates it, $M_{PBH}\propto k^{-2}$. Therefore, more massive PBHs form when small $k$ modes re-enter the horizon, whereas the lighter PBHs form when large $k$ modes re-enter the horizon,  with the amplitude of power spectrum large enough to generate them.
As large $k$ mode leaves the horizon late during inflation and re-enters in the horizon first, this implies that lighter PBHs form early in the radiation era, and the small $k$ modes corresponding to the more massive PBHs enter late and form later in the radiation era.

Further, we can express Eq. (\ref{MBH3}) in terms of the present horizon mass which is given as $
M_0=\frac{4\pi}{3}\rho_{crit}~H_0^{-3}\approx 4.62\times 10^{22} M_{\odot}.
$ 
This is as follows
\begin{align}
M_{PBH}(k)
=&\gamma M_0 \left(\frac{g_{*0}}{g_{*i}}\right)^{1/6}\Omega_{r0}^{1/2}\left(\frac{H_0}{k}\right)^{2}\\
\approx &  ~5\times 10^{15}{\rm{g}}\left(\frac{g_{*0}}{g_{*i}}\right)^{1/6}\left(\frac{10^{15}{\rm{Mpc}}^{-1}}{k}\right)^{2}.
\label{Mpbh}
\end{align}
This relation implies that for an overdense  fluctuation mode with $k\sim 10^{15}$ Mpc$^{-1}$, PBHs of mass $M_{PBH}\sim 5\times10^{15}$ g are formed. 

\subsection{Initial Mass Fraction of Primordial Black Holes}

The present abundances of PBHs are constrained from various observations, which gives
upper limits on their 
initial mass fraction  
defined as 
\begin{equation}
\beta({M_{PBH}})\equiv\frac{\rho_{PBH,i}}{\rho_{{\rm{total}},i}}~.
\label{beta}
\end{equation}
Qualitatively, it is the ratio of the energy density of PBHs of mass $M_{PBH}$ at the time of its formation, $\rho_{PBH,i}$ to the total energy density of the Universe at that epoch, $\rho_{{\rm{total}},i}$. The initial mass fraction of PBHs is a mass dependent quantity.

As the PBH formation is assumed to take place in the radiation dominated era, the total energy density at that epoch is in the radiation, i.e. $\rho_{{\rm{total}},i}=\rho_{ri}$, given in Eq. (\ref{rhoi}), while the energy density of PBHs evolve as $\rho_{{PBH,i}}=\rho_{{PBH,0}}~ a_i^{-3}$. Thus we can write Eq. (\ref{beta}) as 
\begin{equation}
\beta({M_{PBH}})
= \frac{\Omega_{PBH 0}({M_{PBH}})}{\Omega_{r0}} \left(\frac{g_{*i}}{g_{*0}}\right)^{1/3}a_i,
\label{betaa}
\end{equation}
where $\Omega_{PBH 0}({M_{PBH}})=\rho_{{PBH 0}}/\rho_{crit}$ is the density parameter for PBH of mass ${M_{PBH}}$.
On substituting
Eq. (\ref{ai}) in  Eq. (\ref{betaa}) and  then using Eq. (\ref{Mpbh}), we obtain
\begin{align}
\beta({M_{PBH}})=&\frac{\Omega_{PBH 0}({M_{PBH}})}{\Omega_{r0}^{3/4}} \left(\frac{g_{*i}}{g_{*0}}\right)^{1/4}\nonumber \\
\times  & \left(\frac{M_{PBH}}{M_0}\right)^{1/2}\gamma^{-1/2}.
\label{bet}
\end{align}
With this relation, the observational constraints on $\Omega_{PBH 0}$ for a PBH of mass $M_{PBH}$, can be used
to calculate the upper bound on its initial mass fraction $\beta(M_{PBH})$
(see Refs. \cite{Green:1997sz,Carr:2009jm,Josan:2009qn}).
The number of relativistic degrees of freedom at the present, $g_{*0}=3.36$, and $g_{*i}$ denotes at the time of PBH formation in the radiation dominated era. For our supersymmetric model of warm inflation discussed in section \ref{ourWI}, we take $g_{*i} \sim 200$.
The bounds on $\beta(M_{PBH})$ can then be further used to constraint the amplitude of primordial power spectrum, as will be shown in the next subsection. 

\subsection{PBH formation using the Press-Schechter formalism}
\label{PS theory}

Now we discuss the Press-Schechter theory for the formation of a primordial black hole and then show how the primordial power spectrum can be constrained by the PBH abundance calculated using this formalism.  
We consider that the PBH forms by the collapse of a density perturbation generated during  inflation. We assume that the initial gaussian seeds of density perturbations re-enter the horizon during the radiation dominated epoch and the PBH formation takes place in the regions with overdensity above a critical value, $\delta>\delta_c$ where $\delta_c\sim\mathcal{O}$(1) \cite{Carr:1975qj} (see Ref. \cite{Harada:2013epa} for more details). On smoothening the density perturbations using a Gaussian window function, 
the probability distribution for a smoothed density contrast over a radius $R=(aH)^{-1}$ is given as \cite{Liddle},
\begin{equation}
p(\delta(R))=\frac{1}{\sqrt{2\pi}\sigma(R)} \exp\left( \frac{-\delta^2(R)}{2\sigma^2(R)}\right).
\end{equation}
Here $\sigma(R)$ is the mass variance evaluated at the horizon crossing, and is defined as,
\begin{equation}
\sigma^2(R)=\int_{0}^{\infty}\tilde{W}^2(kR) P_\delta(k)\frac{dk}{k}
\label{variance}
\end{equation}
where $P_\delta(k)$ is the matter power spectrum, and $\tilde{W}(kR)$ is the Fourier transform of the window function 
\begin{equation}
\tilde{W}(kR)=\exp(-k^2R^2/2).
\end{equation}
The primordial curvature power spectrum $P_\mathcal{R}(k)$ for the fluctuations generated during the inflation can be related to the density power spectrum $P_\delta(k)$ through the relation,
\begin{equation}
P_\delta(k)=\frac{4(1+w)^2}{(5+3w)^2}\left(\frac{k}{aH}\right)^4 P_\mathcal{R}(k),
\label{PdR}
\end{equation}
where $w$ is the equation of state of the fluid and is equal to 1/3 for a radiation dominated era. Now we carry out the integration in Eq. (\ref{variance}) by substituting
Eq. (\ref{PdR}) into it,  considering 
any parameterization of the primordial power spectrum. 

Using the Press-Schechter theory, the initial mass fraction of a PBH with mass $M_{\rm{PBH}}$ is obtained as \cite{Press:1973iz},
\begin{align}
\beta({M_{PBH}})=&2\int_{\delta_c}^{1} p(\delta(R)) \, d\delta(R) \nonumber \\
=&\frac{2}{\sqrt{2\pi}\sigma(R)}\int_{\delta_c}^{1} \exp\left( \frac{-\delta^2(R)}{2\sigma^2(R)}\right)d\delta(R)\\
=&{\rm{erfc}}\left(\frac{\delta_c}{\sqrt{2}\sigma(R)}\right)
\label{betaM}
\end{align}
where ${\rm{erfc}}$ is the complimentary error function, and we consider $\delta_c=0.5$ in this study. We substitute the mass variance obtained considering any primordial curvature power spectrum from Eq. (\ref{variance}) 
into Eq. (\ref{betaM}).
The expression thus obtained for $\beta({M_{PBH}})$ using Press-Schechter theory is equated to Eq. (\ref{bet}), and constrained using the observational bounds on $\Omega_{PBH 0}({M_{PBH}})$, as argued in the previous subsection. In this way, the primordial power spectrum, and hence inflationary models are constrained from the bounds on abundance of PBHs \cite{Carr:1993aq,Kim:1996hr,Green:1997sz,Josan:2009qn}. For the various mass of PBHs, 
 the upper bound on the amplitude of the primordial power spectrum is obtained to be, $P_\mathcal{R}(k_{PBH}) \sim \mathcal{O}(10^{-2}-10^{-1})$ \cite{Green:2014faa,Sasaki:2018dmp}.

There is a plethora of previous studies which discuss the PBH production from the collapse of large inhomogeneties generated from various inflationary scenarios. Some examples are hybrid inflation model with multiple scalar fields \cite{GarciaBellido:1996qt},  running-mass inflation models \cite{Leach:2000ea,Drees:2011hb,Motohashi:2017kbs}, double inflation model \cite{Saito:2008em}, single field inflation with a broken scale invariance \cite{Bringmann:2001yp}, or by introducing a plateau in the potential \cite{Ivanov:1994pa,Ballesteros:2017fsr},  or running of the spectral index \cite{Drees:2011yz}, etc.
In Ref. \cite{Young:2014ana}, it is shown that  
for a power-law form of the primordial power spectrum, 
in order to form PBHs ( of mass $10^8$ g $<M_{PBH}<10^{10}$ g with $\beta<10^{-20}$ obtained from observations), the spectral index 
has to be $1.26<n_s<1.34$. 
This implies that a blue-tilted power spectrum is required for the formation of PBHs, whereas 
the CMB observations demand that the power spectrum is red-tilted at the large scales, as can be seen from the $n_s$ allowed values from \textit{Planck} 2018 results given later in Eq. (\ref{nsrange}). 
Therefore, it is interesting to obtain some inflationary model which has both these features, and is also consistent with the CMB anisotropy observations.

Further, in Ref. \cite{Green:2018akb} , it is argued that by assuming a running of the spectral index, $\alpha$, 
and running of the running, $\beta$, in the primordial power spectrum, 
one can generate the amplitude of the power spectrum at PBH scales, $P_{\mathcal{R}}(k)\sim\mathcal{O}(10^{-2})$ for some combinations of $\alpha, \beta$.  
But because such a parameterization 
is a non-convergent Taylor series expansion of $n_s(k)$,  the amplitude of $P_{\mathcal{R}}(k)$ 
at large $k$ amplifies by many orders if more terms in the series 
are retained. Therefore, one should look for alternate parameterizations or models of inflation which have a 
 physical origin and interpretation, with the desired features and amplitude of the primordial power spectrum. We present one such model of warm inflation, which can possibly generate the large amplitude of primordial power spectrum required to form PBHs. We first discuss the fundamentals of warm inflation and the primordial power spectrum in the next section, and afterwards study the PBH formation from our warm inflation model.

\section{Theory of warm inflation}
\label{WI}

The basic idea of warm inflation is that the inflaton is sufficiently coupled with other fields during inflation, such that it dissipates its energy into them, that leads to a thermal bath in the Universe during the inflationary phase.
Therefore, the equation of motion of the inflaton $\phi$ slowly rolling down a potential $V(\phi)$ is modified due to an additional dissipation term $\Upsilon\dot{\phi}$ arising from the inflaton coupling to other fields and is given as
\begin{equation}
\ddot{\phi}+(3H+\Upsilon)\dot{\phi}=-V'(\phi).
\label{phieom}
\end{equation}
Here $H$ is the Hubble expansion rate and overdot and $'$ denote the derivative w.r.t. time and $\phi$, respectively.  $\Upsilon$ is called the dissipation coefficient, and it depends on the inflaton field $\phi$ and temperature of the Universe $T$.
Due to the inflaton dissipation,  
there is an energy transfer from the inflaton into the radiation as given by
\begin{equation}
\dot\rho_r+4H\rho_r=\Upsilon{\dot\phi}^2~.
\label{rad}
\end{equation}
We define a dissipation parameter, 
$$Q \equiv \frac{\Upsilon}{3H} $$ 
which is the ratio of the relative strength of inflaton dissipation, compared to the Hubble expansion rate,
and rewrite Eq. (\ref{phieom}) as
\begin{equation}	
\ddot \phi + 3 H( 1+ Q ) \dot\phi + V'(\phi)=0.
\label{inflaton}
\end{equation}
When the dissipation parameter $Q\ll1$, it is the weak dissipative regime of warm inflation, and when $Q\gg1$, it is the strong dissipation regime. For this study, we consider that at the pivot scale ($k_P=0.05$ Mpc$^{-1}$), 
it is the weak dissipation regime of warm inflation. But as $Q$ evolves during inflation, we obtain the strong dissipative regime of warm inflation, by the end of inflation.

In warm inflation, the primordial power spectrum has 
 dominant contributions from the thermal fluctuations of the fields and is 
given in Refs. \cite{Hall:2003zp,Graham:2009bf,Ramos:2013nsa,Bartrum:2013fia} as
\begin{align}
P_\mathcal{R}(k)=&\left(\frac{H_k^2}{2\pi\dot\phi_k}\right)^2\left[1+2n_k+\left(\frac{T_k}{H_k}\right)
\frac{2\sqrt{3}\pi Q_k}{\sqrt{3+4\pi Q_k}}\right]\nonumber \\
 \times & G(Q_k),
\label{power}
\end{align}
where the subscript $k$ represents the time when the  $k^{th}$ mode of cosmological perturbations leaves the horizon during inflation. By the fluctuation-dissipation theorem, the dissipation in the system
sources the thermal fluctuations, which modifies the Langevin equation for the inflaton fluctuations. As a result, the primordial power spectrum gets augmented with an additional term (third in the square bracket) due to the thermal contributions.
Furthermore, the thermal bath of radiation also excites the inflaton 
from its vacuum state 
to a Bose-Einstein distribution $n_k$, given by 
\begin{equation}
n_k=\frac{1}{\exp(\frac{k/a_k}{T_k}) -1} ~,
\label{BE}
\nonumber
\end{equation}
which alters the primordial power spectrum. 
Additionally, as the dissipation coefficient $\Upsilon$ depends on the temperature, therefore 
the perturbations in the radiation
 can also couple to the inflaton perturbations and lead to a growth in the primordial power spectrum \cite{Graham:2009bf}. This growth factor $G(Q_k)$ depends on the form of $\Upsilon$ and is obtained numerically, as shown in Ref. \cite{Graham:2009bf}. For a dissipation coefficient with a cubic dependence on temperature $\Upsilon\propto T^3$, the growth factor   \cite{Benetti:2016jhf} 
$$ G(Q_k)_{cubic}= 1+4.981 \,Q_k^{1.946} +0.127 \,Q_k^{4.330}.$$
In the weak dissipation regime for small $Q$, the growth factor does not enhance the power spectrum much. But for $Q\gg1$ in the strong dissipation regime, the power spectrum is  significantly enhanced due to the growth factor. In the strong dissipative regime, the shear effects in the radiation also becomes important at the perturbative order, and cause damping of the power spectrum \cite{BasteroGil:2011xd}. However, if the self-coupling in the radiation are strong enough, then the effect of shear can be reduced, and a net growth of the primordial power spectrum is obtained (see Ref. \cite{BasteroGil:2011xd} for details).

\section{Warm Inflation Model and Formation of PBHs}
\label{ourWI}

Now we discuss the formation of PBHs from our model of warm inflation. 
We consider a supersymmetric model of warm inflation with a two stage decay mechanism, where the inflaton couples to  intermediate $X$ superfields, which decay into $Y$ superfields, which thermalise and form a radiation bath.
 In this study, we consider a quartic potential of warm inflation
$(V(\phi)=\lambda\phi^4)$ with a 
dissipation coefficient given by \cite{Moss:2006gt},
\begin{equation}
\Upsilon(\phi,T)=C_\phi \frac {T^3}{\phi^2}.
\label{upsiloncubic}
\end{equation}
This kind of dissipation term emerges in the low temperature regime of warm inflation, when the inflaton couples to the heavy ($m_X\gg T$) bosonic components of the $X$ fields  \cite{Moss:2006gt,BasteroGil:2010pb,BasteroGil:2012cm}.
The constant $C_{\phi}$ depends on the couplings and the multiplicities of $X$ and $Y$ as 
$C_\phi=\frac{1}{4}\alpha N_X, $ 
where $\alpha=h^2 \frac{N_Y}{4\pi}<1$, $h$ is the coupling between $X$ and $Y$, and $N_{X, Y}$ are the multiplicities of the $X$ and $Y$ fields \cite{BasteroGil:2012cm,Bartrum:2013fia}.

The motivation for considering this warm inflationary model is that 
it is consistent with the CMB observations for some parameter space of the model parameters. 
Also, the tensor-to-scalar ratio prediction for this model is within the sensitivity of the next generation CMB polarisation experiments and therefore can be tested in the near future.
Furthermore, we shall see that the primordial power spectrum for this model of warm inflation has features required for PBH generation. These features arise due to intrinsic properties of the inflaton-radiation system and therefore are interesting to study.

In our earlier work in Ref. \cite{Arya:2017zlb}, we  parameterized the primordial power spectrum for this model in terms of two model parameters - the inflaton self coupling, $\lambda$, and the dissipation parameter at the pivot scale, $Q_P$, and  estimated them using large scale CMB observations. 
By using the same parameterization, we study the formation of small scale PBHs for our warm inflationary model considering different values of $Q_P$ in this work. For each  $Q_P$ value, we consider $\lambda$ such that the primordial power spectrum is normalised at the pivot scale as $P_\mathcal{R}(k_P)=2.1\times 10^{-9}.$

\section{Analysis and Discussion}
\label{discussion}
In this section, we analyze the results of our study about the formation of PBHs from warm inflation model. 
Firstly, we  
plot the primordial power spectrum for our warm inflation model as a function of $k$, following the parameterization in our previous study \cite{Arya:2017zlb}. 
In order to study the role of the dissipation parameter, $Q_P$ on the formation of PBHs,
we consider different values of $Q_P$ to plot the primordial power spectrum. Then we discuss the range of $Q_P$ values allowed from the  $n_s$ observations. 
Further, we discuss the effects of higher dissipation on the initial mass fraction 
and the mass of the generated PBHs.
Finally, we remark on the implications of the calculated initial mass fraction for tiny PBHs formed from our warm inflation model, and the possibilty of PBH relics as dark matter.

\subsection{Primordial power spectrum for our warm inflation model }

\begin{figure}[tbp]
	\centering
\includegraphics[width=2.7in,height=2.4in]{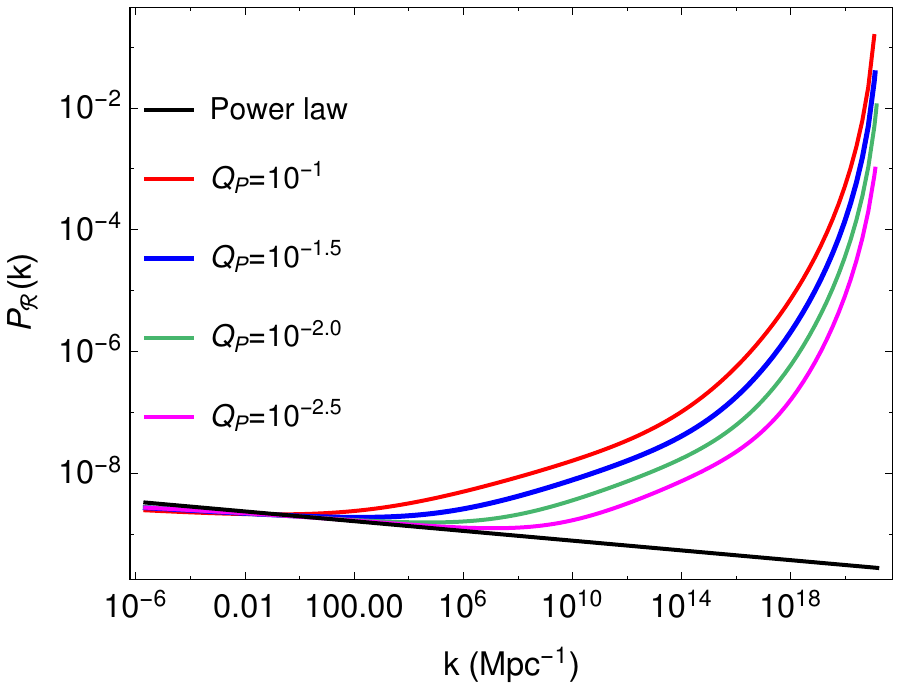}
\caption{Plot of the primordial power spectrum $P_{\mathcal{R}}(k)$ versus $k$ for our warm inflation model with different values of $Q_P$. 
Here Black line represents the standard power law parameterization considered in cold inflation.}
\label{pkplot}
\end{figure}

 We first plot the primordial curvature power spectrum as a function of the comoving wavenumber $k$ in Fig. \ref{pkplot}. 
We consider that the number of efolds when the pivot scale ($k_P=0.05$ Mpc $^{-1}$) leaves the horizon, $N_P=60.$ (In our notation, at the end of inflation, $N_e=0$.)
As already mentioned, PBHs provide a probe for a huge range of small scale modes. Here we consider only those $k$ modes that leave the horizon near the end of inflation, and can form PBHs when they re-enter in the radiation era.

In order to produce a significant number of PBHs, the amplitude of $P_{\mathcal{R}}(k)$ needs to be $\mathcal{O}(10^{-2})$ \cite{Green:2014faa}. We consider various cases of inflation with different values of the dissipation parameter at the pivot scale, $Q_P=10^{-1}, 10^{-1.5},$ $ 10^{-2},$ and $10^{-2.5}$ (weak dissipation regime when the CMB scales exit the horizon) to plot Fig. \ref{pkplot}. As can be seen, for some $Q_P$ values, a large amplitude of $P_{\mathcal{R}}(k)$ is achieved near the end of inflation at $k\sim 10^{21}$ Mpc$^{-1}$. 
These small scale modes leave the horizon when inflation is near its end, and then re-enter in the horizon during radiation dominated era. When they re-enter, these overdense perturbations collapse to form the primordial black holes, as discussed in Section \ref{PBH form}.
For the strong dissipation regime $Q_P>1$ as well, the amplitude of $P_{\mathcal{R}}(k)$ would  be $\mathcal{O}(10^{-2})$ and higher, but those cases are not of interest, for the reason we discuss later.

We also plot the power law power spectrum parameterization, considered in cold inflation (without running of $n_s$) (Black line) in Fig. \ref{pkplot} for comparison. It can be seen that for a power law power spectrum, the amplitude of $P_{\mathcal{R}}(k)$ can never reach the large value $\mathcal{O}(10^{-2})$, as the spectrum is red-tilted ($n_s<1$), and therefore no PBHs can be produced for such a form of $P_{\mathcal{R}}(k)$.
But for our model of  warm inflation, we find that the power spectrum changes to blue-tilted ($n_s>1$) at the PBH scales and therefore 
PBH formation can take place for some range of model parameters.

Now we discuss the effects of the inflaton dissipation during warm inflation to the primordial power spectrum.
 It can be seen from Fig. \ref{pkplot} that at the PBH scales (large $k$), for a large dissipation parameter $Q_P$, the amplitude of the primordial power spectrum $P_{\mathcal{R}}(k)$ is larger as compared to the small dissipation case. This implies that for a large $Q_P$, the amplitude of $P_{\mathcal{R}}(k)$ is enhanced to $\mathcal{O}(10^{-2})$ at a comparatively smaller $k$, 
and all the larger $k$ modes leaving the horizon further 
are sufficiently overdense to form PBHs. From the plot, it is also seen that 
for $Q_P<10^{-2}$, the amplitude of primordial power spectrum is 
not sufficient to generate PBHs.
Therefore, we limit our study of PBH formation till $Q_P=10^{-2}$.

\subsection{Relevant range of the dissipation parameter consistent with CMB }
\begin{figure}[]
	\centering
\includegraphics[width=2.5in,height=1.75in]{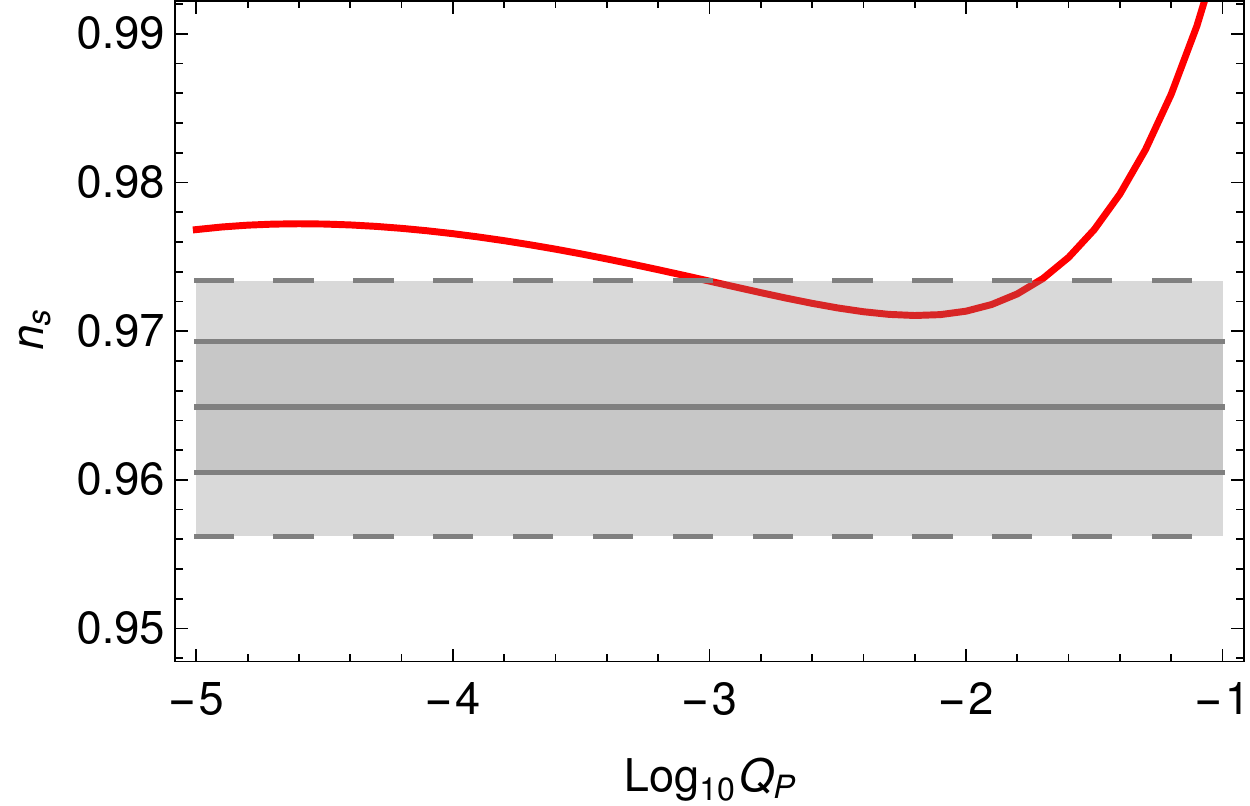}
\caption{Plot of the spectral index at the pivot scale, $n_s$ versus $\log_{10}Q_P$. The colored band signifies the results of CMB observations from Planck 2018 for the allowed range of $n_s$ within $68\%$ and $95\%$ C.L..}
\label{nsplot}
\end{figure}

Next, we plot the spectral index of the primordial power spectrum at the pivot scale, $n_s$ for the different values of the dissipation parameter, $Q_P$ in Fig. \ref{nsplot}. For reference, we also plot the allowed range of $n_s$ values in the $68\%$ and $95\%$ C.L. from the recent Planck 2018 results for TT,TE,EE + lowE dataset as given in Ref. \cite{Aghanim:2018eyx}, 
\begin{align}
& n_s=0.9649\pm 0.0044 ~~(68\% ~ {\rm{C.L.}}), ~~~ {\rm{and}} \nonumber \\ & n_s=0.9649^{+0.0085}_{-0.0087} ~~(95\% ~ {\rm{C.L.}}).
\label{nsrange}
\end{align}

From Fig. \ref{nsplot}, it can be seen that only a small range of $Q_P$ values are consistent with the CMB observations, and for $Q_P>1$, the value of $n_s$ is much higher $(>1)$ than the allowed values. Therefore we do not consider $Q_P>10^{-1.7}$, despite the fact that 
the amplitude of the primordial power spectrum at PBH scales, $P_\mathcal{R}(k_{PBH})\sim \mathcal{O}(10^{-2})$, which is sufficient to form PBHs.

\subsection{Initial mass fraction and mass of the PBH formed}
Now, we are equipped with the knowledge that for a certain range of $Q_P$ values, the amplitude of the primordial power spectrum at the PBH scales, $P_\mathcal{R}(k_{PBH})\sim \mathcal{O}(10^{-2})$, is sufficient enough to  generate PBHs.  For each scenario of inflation, represented by the different values of $Q_P$, we first calculate the mass variance  by substituting the warm inflation power spectrum from Eqs. (\ref{power}), and (\ref{PdR}) in Eq. (\ref{variance}). The mass variance, $\sigma^2(R)$ is obtained as a function of $R=(aH)^{-1}=k^{-1}$ at the horizon crossing, which can be then related to $M_{PBH}$ through Eq. (\ref{Mpbh}). After that, we substitute the obtained mass variance in Eq. (\ref{betaM}) and numerically calculate the initial mass fraction for the PBHs as a function of its mass, $\beta(M)$. 

\begin{figure}[tbp]
\centering
\includegraphics[width=2.75in,height=1.8in]{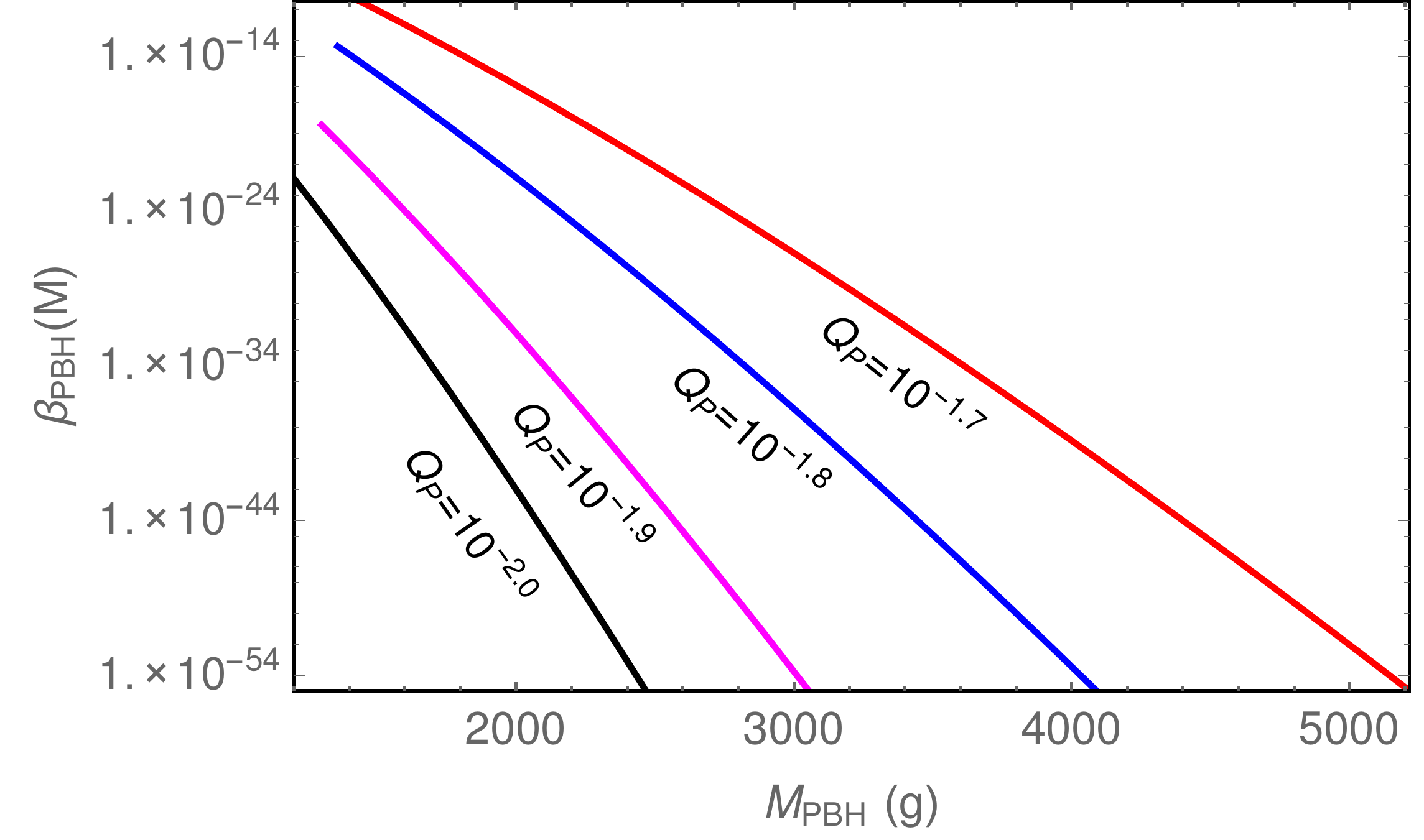}	\caption{Plot of the initial mass fraction of the generated PBH, $\beta(M_{PBH})$, versus its mass $M_{PBH}$ (g). Note that such tiny mass PBHs would have evaporated by today.}
\label{betaplot1}
\end{figure}

 Finally, we plot the obtained inital mass fraction $\beta(M)$  of the generated PBH versus the mass of the PBH in Fig. \ref{betaplot1} 
for the cases when $Q_P=10^{-1.7}, 10^{-1.8},10^{-1.9},$ and $10^{-2}.$
As shown in Fig. \ref{pkplot}, for $Q_P<10^{-2}$, the amplitude of $P_\mathcal{R}(k)$ is not sufficient to form PBHs, and hence is not considered in this study.

From Fig. \ref{betaplot1}, we can see that the mass of PBHs generated from our warm inflation model is of the order $M_{PBH}\sim 10^3$ g.
From the plots, we infer 
that a large dissipation during inflation leads to a comparatively more massive PBH formation, whereas  small dissipation produces small mass PBHs. The reason for this is that, as shown in Fig. \ref{pkplot}, for a larger dissipation, the desired amplitude of the primordial power spectrum $P_\mathcal{R}(k_{PBH})\sim \mathcal{O}(10^{-2})$, is achieved at a comparatively smaller $k$. As the mass of the generated PBH is proportional to  $k^{-2}$, this implies that PBHs formed by small $k$ overdense modes are more massive, compared to the large $k$ overdense modes.

\subsection{Constraints on the abundance of PBHs formed}

 As the range of $Q_P$ values relevant for PBH study is very small, 
there is a very narrow range of $k$ modes available for the PBH generation, and consequently a narrow range of mass of the generated PBHs.  The order of mass of the PBH formed from our warm inflation model, $M_{PBH}\sim 10^3$ g. Such  tiny mass PBHs have an extremely short lifetime $\sim 10^{-19}$ sec and would have evaporated by now into Hawking radiation \cite{Hawking:1974rv, Hawking:1974sw}. The abundance of such tiny mass PBHs is not strictly bounded. However, there are certain bounds from the 
PBH evaporation leading to the generation of stable massive particles (Supersymmetric LSP) \cite{Green:1999yh} or long lived decaying particles (eg. gravitino, modulii) \cite{ Lemoine:2000sq,Khlopov:2004tn},
and their relic abundance,
which can be used to put constraints on PBH initial abundance \cite{Green:2014faa,Josan:2009qn,Carr:2009jm}.
To an order of magnitude, for $M_{PBH}\sim10^3$ g, the upper bound on $\beta(M_{PBH})<10^{-14}.$  For various scenarios (different $Q_P$) of our warm inflation model, we find that the calculated initial abundance of PBHs is in accordance with the observational limits for $Q_P=10^{-1.8}, 10^{-1.9}$, and $10^{-2}$. But for the case with $Q_P=10^{-1.7}$, the theoretical estimate of the initial mass fraction is higher than the 
above mentioned evaporation constraints, which implies that this case is inconsistent with the PBH bounds, and should be ruled out.

\subsection{Generated PBHs as constituent of dark matter}
It is argued that Primordial Black Hole evaporation could leave a stable relic of Planck mass, which can contribute to the Dark matter \cite{MacGibbon:1987my,Chen:2003bn}. In order that the Planck mass relics do not overclose the Universe today, the present density of Planck mass relics should be less than the present cold dark matter density. 
To an order of magnitude, the constraints on the initial mass fraction of  PBH of mass $M_{PBH}\sim 10^3$ g, as calculated in Ref. \cite{Carr:1994ar} is given to be 
$\beta(10^3 g)<10^{-16}.
$ (Also see, Refs. \cite{Carr:2005zd,Carr:2009jm,Josan:2009qn}) 
We find that in our warm inflation models with $Q_P=10^{-1.9},$ and $Q_P=10^{-2}$, the initial mass fraction lies within the above estimated limits, and therefore the possibility that PBH remnants form DM, remains valid for these cases.
But as the Planck mass relics are very small, it is extremely difficult or nearly impossible to observationally detect them non-gravitationally.

\section{Summary}
\label{Summary}
Primordial Black Holes are a remarkable probe to the physics of the early Universe. They provide us an opportunity to investigate a huge range of scales of perturbations generated during the inflationary phase. In this study, we consider one model of warm inflation scenario, and discuss the PBH formation by the collapse of large inhomogeneties generated during it.
Warm inflation is an alternate description of inflation in which the radiation production takes place concurrent to the inflationary phase. The inflaton couples and dissipates into the other fields during inflation, which creates a thermal bath of particles during inflation. The inflaton dissipation characterized by a parameter $Q_P$ contributes to the primordial power spectrum, and enhances the amplitude of the primordial power spectrum at the small scales to $\mathcal{O}(10^{-2})$, required to generate PBHs.  

We find that for some parameter range of our model, PBHs can be generated with a significant abundance.
We consider those cases with values of the dissipation parameter at the pivot scale as, $Q_P=10^{-1.7}, 10^{-1.8}, 10^{-9}, 10^{-2}$. 
We calculate the initial mass fraction and the mass of the generated PBHs for these values of the dissipation parameter.
We obtain that our model of warm inflation can produce a significant abundance of  PBHs with mass, $M_{PBH}\sim10^3$ g. Such tiny mass PBHs have a very short lifetime of $10^{-19}$ sec, and would have evaporated into Hawking radiation. 
By an order of estimate, we found that for the cases with $Q_P= 10^{-1.8}, 10^{-9}, 10^{-2}$, the obtained initial mass fraction is in accordance with the upper limit obtained from the abundance of stable and long lived decaying particles produced by evaporating PBHs. The case with $Q_P= 10^{-1.7}$ overproduces PBHs, which is inconsistent with the upper bounds on $\beta$, and hence should be ruled out.

Furthermore, it is also argued that PBH evaporation ceases when PBH mass gets close to the Planck mass, and such Planck mass relics can thus constitute the present dark matter. The present density of the Planck mass relics should be less than the cold dark matter density, so that it does not overclose the Universe today. This gives a rough bound on the PBH initial mass fraction for a PBH of mass $10^3$ g of an order, $\beta(10^3 g)< 10^{-16}.$ For our warm inflation models with $Q_P=10^{-1.9},$ and $Q_P=10^{-2},$ we find that the calculated initial mass fraction lies within the limits, and hence the possibility to form DM remains valid. However, Planck mass relics are extremely tiny and almost impossible to detect by non-gravitational measures.

\begin{acknowledgements}
The author would like to thank Prof. Raghavan Rangarajan and Prof. Namit Mahajan for their valuable suggestions and comments. RA is also thankful to her colleague Arvind Kumar Mishra for the fruitful discussions and suggestions during the preparation of the manuscript.
\end{acknowledgements}

\bibliographystyle{utphys}
\bibliography{ref}

\end{document}